\documentclass[preprint,12pt]{elsarticle}

\usepackage[margin=2.1cm]{geometry}
\usepackage{amsmath,amssymb}
\usepackage{graphicx}
\usepackage{booktabs}
\usepackage{bm}
\usepackage{enumerate}
\usepackage{xcolor}
\definecolor{linkblue}{rgb}{0.0,0.2,0.5}
\usepackage[colorlinks=true,linkcolor=linkblue,citecolor=linkblue,urlcolor=linkblue]{hyperref}
\usepackage{xspace}
\usepackage{orcidlink}

\graphicspath{{figs/}}

\newcommand{\ulp}{\text{ulp}\xspace}
\newcommand{\vp}{\bm{p}}
\newcommand{\vbeta}{\bm{\beta}}
\newcommand{\vu}{\bm{u}}
\newcommand{\nhat}{\hat{\bm{n}}}
\newcommand{\tol}{\tau}

\def\pythia{\textsc{Pythia}\xspace}
\def\herwig{\textsc{Herwig}\xspace}
\def\sherpa{\textsc{Sherpa}\xspace}
\def\rivet{\textsc{Rivet}\xspace}
\def\fastjet{\textsc{FastJet}\xspace}
\def\root{\textsc{Root}\xspace}
\def\clhep{\textsc{Clhep}\xspace}
\def\thepeg{\textsc{ThePEG}\xspace}
\def\vect{\textsc{Vector}\xspace}
\def\smashlib{SMASH\xspace}
\def\corsika{\textsc{Corsika}\xspace}

\newcommand{\figref}[1]{fig.~\ref{#1}\xspace}
\newcommand{\secref}[1]{sec.~\ref{#1}\xspace}
\renewcommand{\eqref}[1]{eq.~(\ref{#1})\xspace}

\newcommand{\Secref}[1]{Section~\ref{#1}\xspace}

\journal{Computer Physics Communications}

\begin{document}

\begin{frontmatter}

\title{Lossy Lorentz boosts at ultra-relativistic energies\\ \normalsize{Numerical study and source survey of high-energy physics software}}

\author{Christian Bierlich \orcidlink{0000-0002-3978-6085}}
\ead{christian.bierlich@fysik.lu.se}
\address{Department of Physics, Lund University, S\"olvegatan 14A, SE-223 62 Lund, Sweden}

\begin{abstract}
\noindent
Most high-energy-physics software packages implement Lorentz boosts for themselves. At sufficiently large boosts, these routines can return finite, plausible-looking four-vectors whose invariant masses and rest-frame properties are no longer accurate. We survey implementations in widely used collider and air-shower software and compare representative algorithms in three numerical tests.

We find that the main problem is usually information discarded by the representation, rather than the arithmetic of the boost formula. Information can be lost when a boost is specified only by its velocity, and again when the result is stored only through Cartesian components. Richer boost input prevents the first loss, while retaining light-cone coordinates prevents the second. Ordinary hard processes at present colliders are safe in double precision; the relevant regimes are low-mass systems, highly boosted auxiliary frames, single-precision particle records and air-shower simulations. Extreme scale hierarchies in parton-shower accuracy tests and ultra-heavy particle decays provide still more demanding cases.
\end{abstract}


\end{frontmatter}

\section{Introduction}
\label{sec:intro}

The Lorentz boost is elementary mathematics implemented in a dozen lines of
code, and every major high-energy physics software package carries its own copy. Those
copies are not equivalent numerically. Knowledge of their floating-point
behaviour is fragmentary and scattered: the \vect package
documentation asserts, without published measurement, that boosting by a
four-momentum is ``more numerically stable'' than converting to a velocity
first~\cite{Chopra:2025faa}; the plasma-simulation literature knows that evaluating
$\gamma=\sqrt{1+u^2}$ from the proper velocity avoids the cancellation
that afflicts $1/\sqrt{1-\beta^{2}}$~\cite{Vay:2008,Schmitz:2026}; and the
\textsc{Corsika}~8 framework names accuracy ``from non-relativistic to
ultra-relativistic boosts'' as a design requirement~\cite{Engel:2018akg} without
mapping which formulas meet it. General-purpose Monte Carlo event generator codes for high-energy physics, such as \pythia~\cite{Bierlich:2022pfr}, \herwig~\cite{Bewick:2023tfi} and \sherpa~\cite{Sherpa:2024mfk}, as well as more specialized codes such as \smashlib~\cite{SMASH:2016zqf}, also carry their own implementations which in some cases predate the routine use of double-precision arithmetic. No systematic cross-package source survey of these interfaces and their failure modes exists.

The question has become practical. General-purpose event generators are now
both embedded in air-shower simulation~\cite{CORSIKA:2025ops}, and are even starting
to implement the full air-shower chain from initiator to ground themselves \cite{Lonnblad:2025ivw}.
Analysis frameworks such as \root~\cite{Brun:1997pa} are already routinely used for such physics analyses, and ongoing tuning efforts for air-shower simulations~\cite{Albrecht:2025kbb} rely on a software stack mixing traditional air-shower simulators and analysis software traditionally
belonging in high-energy physics, such as \rivet~\cite{Bierlich:2024vqo}. We furthermore note that this study is directly motivated by such an error being uncovered during development of \pythia, only through tracing a resulting floating-point exception, and a fair bit of luck.

The relevant boost is not fixed by the collision energy alone. In air showers, one reaches initiator proton energies of up to
$10^{20}\,$eV, corresponding to $\gamma\sim 10^{11}$. Hard systems at
colliders typically have $\gamma$ between unity and a few hundred; electrons
and low-mass subsystems can reach $\gamma\sim10^5$.
Tests of the logarithmic accuracy of parton showers probe a different extreme:
nearly massless intermediate systems generate very large boosts even at fixed
overall energy~\cite{Dasgupta:2020fwr,Herren:2022jej,Preuss:2024vyu}.

The central observation of this paper is that a floating-point exception following a catastrophic cancellation is the \emph{fortunate} outcome. Cartesian boost algorithms, including those which receive the Lorentz factor
explicitly, return finite, plausible-looking four-vectors whose invariant
masses and rest-frame angles are wrong over a range of Lorentz factors that
begins well before anything fails visibly. None of the usual defences --- a
check for non-finite numbers, a trap on floating-point exceptions, a guard on
$m^2<0$ or on $\beta^2\geq1$ --- fires in that range. The same range is invisible
to the natural component-wise accuracy test of a boost. We call these
\emph{invisible failures}. A failure is \emph{visible} when the calculation
itself exposes it through, for example, a non-finite result, an exception or an
explicit warning, without comparison to an independently known answer.

The obvious way to explain such failures --- that some boost formula is
rounded badly --- is mostly wrong. Two logically separate error sources must be
distinguished:

\begin{enumerate}
 \item How much information is lost by the \emph{representation} of the numbers going into, or coming out of, the boost? We refer to this error source as \textit{representation loss} below.
 \item How much error does the algorithm consuming those doubles add on top? We refer to this error source as \textit{algorithmic error} below.
\end{enumerate}

\Secref{sec:error-sources} defines this decomposition, and shows that the
failures observed in practice are representation loss in two places: in the
velocity $\vbeta$ given as input, and in the Cartesian components
$(E,p_x,p_y,p_z)$ in which the boosted vector is returned. The first can be
avoided by passing richer input. The second cannot be avoided by any boost
formula that returns Cartesian components, because those components no longer
contain the invariants of the boosted vector.

The paper therefore has three parts. After introducing the problem, \secref{sec:algos} and \secref{sec:libs} introduce the compared boost algorithms and the surveyed library
implementations. \Secref{sec:bench} and \secref{sec:results} demonstrate both
invisible and visible failures in three simple, generator-independent tests:

\begin{enumerate}
	\item Component-wise accuracy of random boosts of random four-vectors.
	\item The reconstructed mass of a boosted resonance.
		\item The reconstructed rest-frame decay angle after a two-body decay is boosted through chains along different axes, and back.
\end{enumerate}

Finally, the representation-level remedy is shown to work: a vector which
retains light-cone coordinates along the boost axis keeps its
invariants to the precision of the rest frame at any Lorentz factor tested, a
chain of boosts composed into a single boost before it is applied does the same
unless the chain nearly cancels, and we discuss how a check for cancellation in
Cartesian operations could make the remaining failures visible.

\section{Relevant physical scales}
\label{sec:physical-scales}

The relevant Lorentz $\gamma$-factor belongs to the particle, subsystem or frame being
transformed, not to the collider as a whole. For a massive object of energy
$E$ and invariant mass $M$,
\begin{equation}
  \gamma = \frac{E}{M}.
\end{equation}
The same energy therefore gives very different boosts for e.g.~a pion, a proton and
a heavy resonance or jet system. In double precision, with
$\varepsilon=2^{-52}$, the relative loss associated with reconstructing a
small invariant from Cartesian components grows approximately as
$\gamma^2\varepsilon$, as will be shown in \secref{sec:error-sources}. We use this
scale below to compare physical boosts with experimental precision.
Table~\ref{tab:physical-gammas} gives representative values for an object
carrying the quoted energy. The purpose of this section is to make clear, based on concrete physics cases, which $\gamma$-values are sensible to benchmark in the following.

\begin{table}[t]
  \centering
  \small
  \begin{tabular}{lcccc}
    \toprule
    Energy scale & $E$ &
    $\gamma$ at $M=m_\pi$ &
    $\gamma$ at $M=m_p$ &
    $\gamma$ at $M=100~\mathrm{GeV}$ \\
    \midrule
    LEP/FCC-ee $Z$-pole beam energy  & $\sim 50~\mathrm{GeV}$  &
    $\sim 10^{3}$  & $\sim 10^{2}$  & --- \\
    LHC beam energy                   & $\sim 7~\mathrm{TeV}$   &
    $\sim 10^{5}$  & $\sim 10^{4}$  & $\sim 10^{2}$ \\
    FCC-hh beam energy                & $\sim 50~\mathrm{TeV}$  &
    $\sim 10^{6}$  & $\sim 10^{5}$  & $\sim 10^{3}$ \\
    Cosmic-ray knee                & $\sim 5~\mathrm{PeV}$   &
    $\sim 10^{8}$  & $\sim 10^{7}$  & $\sim 10^{5}$ \\
    Cosmic-ray ankle               & $\sim 5~\mathrm{EeV}$   &
    $\sim 10^{11}$ & $\sim 10^{10}$ & $\sim 10^{8}$ \\
    Ultra-high-energy cosmic ray   & $\sim 100~\mathrm{EeV}$ &
    $\sim 10^{12}$ & $\sim 10^{11}$ & $\sim 10^{9}$ \\
    \bottomrule
  \end{tabular}
  \caption{Approximate Lorentz factors $\gamma=E/M$ for representative
  energies and invariant masses. Each entry refers to one particle or
  subsystem carrying the quoted energy.}
  \label{tab:physical-gammas}
\end{table}

\subsection{Hard systems and precision resonances}

Heavy particles and hard subsystems at present and near-future colliders
typically have modest Lorentz factors. An object of mass $100~\mathrm{GeV}$
carrying the full $7~\mathrm{TeV}$ LHC proton energy would have
$\gamma\simeq70$, and typical values are smaller. At FCC-ee, for example,
Higgsstrahlung at
$\sqrt{s}=240~\mathrm{GeV}$ gives
\begin{equation}
  \gamma_H \simeq 1.08,
  \qquad
  \gamma_Z \simeq 1.15.
\end{equation}
For these Lorentz factors, $\gamma^2\varepsilon$ remains of order $\varepsilon$, i.e.\  machine precision.

Precision resonance measurements give a second useful scale for practical applications.
Table~\ref{tab:resonance-precision} gives the Lorentz factor at which
$\gamma^2\varepsilon$ reaches the current experimental uncertainty on the mass
of three well-measured resonances. These are order-of-magnitude scales; the
benchmark below contains order-one kinematic factors, so its crossings need not
occur at exactly the same Lorentz factors.

\begin{table}[t]
  \centering
  \begin{tabular}{lcccc}
    \toprule
    Resonance & Mass & Mass uncertainty &
    $\delta m/m$ &
    $\gamma$ at $\gamma^2\varepsilon=\delta m/m$ \\
    \midrule
    $J/\psi$       & $3.096900~\mathrm{GeV}$ &
    $6~\mathrm{keV}$ &
    $1.9\times10^{-6}$ & $\sim 10^{5}$ \\
    $\Upsilon(1S)$ & $9.46040~\mathrm{GeV}$ &
    $100~\mathrm{keV}$ &
    $1.1\times10^{-5}$ & $\sim 2\times10^{5}$ \\
    $Z$             & $91.1880~\mathrm{GeV}$ &
    $2.0~\mathrm{MeV}$ &
    $2.2\times10^{-5}$ & $\sim 3\times10^{5}$ \\
    \bottomrule
  \end{tabular}
  \caption{Current resonance masses and uncertainties from the Review of
  Particle Physics~\cite{ParticleDataGroup:2024cfk}. The final column translates
  the relative mass uncertainty into the scale $\gamma^2\varepsilon$ studied
  below.}
  \label{tab:resonance-precision}
\end{table}

Assigning the full $7~\mathrm{TeV}$ LHC proton energy to a single resonance
gives the conservative upper bounds
\begin{equation}
  \gamma_{J/\psi}\simeq2\times10^{3},
  \qquad
  \gamma_{\Upsilon}\simeq7\times10^{2}.
\end{equation}
The corresponding values of $\gamma^2\varepsilon$ are approximately three
orders of magnitude below the current relative $J/\psi$ mass uncertainty and
five orders below that of the $\Upsilon(1S)$.

FCC-ee projects a statistical uncertainty of about $4~\mathrm{keV}$ on
$m_Z$, with an uncertainty of about $100~\mathrm{keV}$ from beam-energy
calibration~\cite{AlcarazMaestre:2021ssl,FCC:2025lpp}. Even at this precision,
the relevant numerical scale is close to machine precision because the
produced boson has $\gamma_Z\simeq1$.

\subsection{Low-mass systems}

The situation changes for particles and subsystems with small invariant
masses. At either LEP or FCC-ee running at the $Z$ pole, a beam electron has
\begin{equation}
  \gamma_e
  =
  \frac{45.6~\mathrm{GeV}}{m_e}
  \simeq 9\times10^{4},
\end{equation}
while at $\sqrt{s}=240~\mathrm{GeV}$ the beam electrons have
$\gamma_e\simeq2.3\times10^{5}$. The corresponding values of
$\gamma_e^2\varepsilon$ are $2\times10^{-6}$ and $1.2\times10^{-5}$.

Low-mass composite systems reach the same regime. An $e^+e^-$ pair produced
near threshold and carrying $100~\mathrm{GeV}$ has
$\gamma_{ee}\simeq E_{ee}/(2m_e)\simeq10^5$. At this boost, reconstructing its
small invariant from Cartesian energy and momentum components can amplify
round-off to a few parts in $10^6$. Low-mass systems can therefore enter the
numerically sensitive regime at ordinary collider energies, even when the
hard system containing them does not.

This sensitivity is known in precision Bhabha calculations, where BBBREM
rearranges expressions to retain contributions of relative order $m_e^2/E^2$
when transforming between its auxiliary and laboratory
frames~\cite{Kleiss:1994wq}.

\subsection{Nearly null systems}

Large Lorentz factors also arise when a subsystem has an exceptionally small
invariant mass. For two massless particles of equal energy $E$, separated by
an angle $\theta$,
\begin{equation}
 M^2=2E^2(1-\cos\theta),\qquad
 \gamma_{12}=\frac{2E}{M}=\frac{1}{\sin(\theta/2)}
 \simeq\frac{2}{\theta}.
 \label{eq:nearly-null-gamma}
\end{equation}
The boost of the pair therefore grows without bound in the collinear limit,
even at fixed energy.

This limit is used deliberately in tests of the logarithmic accuracy of parton
showers. The PanScales and Alaric constructions probe emissions across extreme
soft and collinear hierarchies~\cite{Dasgupta:2020fwr,Herren:2022jej}. In the
corresponding Apollo tests, the momenta are constructed directly in the event
centre-of-mass frame, avoiding boosts into antenna-end systems whose invariant
mass vanishes~\cite{Preuss:2024vyu}. These examples concern numerical limits of used
algorithms rather than high-energy behaviour of (simulated) collider events.

\subsection{Fixed-target and cosmic-ray kinematics}

In fixed-target collisions, the complete collision and its individual
particles probe different boost scales. A $7~\mathrm{TeV}$ proton incident on
a stationary proton gives
\begin{equation}
  \gamma_{\mathrm{cm}}
  \simeq
  \sqrt{\frac{E_{\mathrm{beam}}}{2m_p}}
  \simeq 60 .
\end{equation}
An $\Upsilon(1S)$ produced near rest in this centre-of-mass frame therefore
enters the laboratory with $\gamma_\Upsilon$ of the same order. Forward light
hadrons reach much larger Lorentz factors through their smaller masses.

Cosmic-ray simulations extend the range further. A $1~\mathrm{PeV}$ pion has
$\gamma\simeq7\times10^{6}$, and a $100~\mathrm{EeV}$ proton has
$\gamma\simeq10^{11}$. The centre-of-mass frame of the latter colliding with a
stationary nucleon has $\gamma_{\mathrm{cm}}\sim2\times10^{5}$. Air-shower
simulations construct, decay and transform particles across this full
hierarchy of frames~\cite{Engel:2018akg,CORSIKA:2025ops}.

Ultra-heavy particle decays give a related hierarchy. Calculations of dark
matter decay spectra extend from the electroweak scale to the Planck
scale~\cite{Bauer:2020jay}. In a two-body decay where a $Z$ receives half the
parent mass as energy, parent masses of $1~\mathrm{PeV}$, $2~\mathrm{EeV}$ and
$10^{19}~\mathrm{GeV}$ give
\begin{equation}
 \gamma_Z\simeq5\times10^3,\qquad 10^7,\qquad 5\times10^{16},
 \label{eq:heavy-parent-gammas}
\end{equation}
respectively. These three values are included explicitly in the resonance
benchmark below.

Hard systems at present colliders therefore populate $\gamma$ values from
unity to a few hundred; low-mass collider subsystems readily reach
$\gamma\sim10^{5}$; and cosmic-ray particles extend the range to
$\gamma\sim10^{12}$. Ultra-heavy decays and formal shower tests extend the
hierarchy further. These are the scales used in the numerical tests below.

\section{Two sources of error}
\label{sec:error-sources}
We describe in detail the two sources of error, representation loss and algorithmic error, mentioned in the introduction. The former is split into representation loss in the input (a) and the output (c) below.

\subsection*{(a) Representation loss in the boost input}
A pure boost is fixed by a rapidity $\eta$ and a direction $\nhat$, from
which
\begin{equation}
\label{eq:definitions}
\gamma=\cosh\eta,\qquad \beta=\tanh\eta,\qquad
u=|\vu|=\sinh\eta,\qquad \vu=\gamma\vbeta.
\end{equation}
We use $\eta$ for the boost rapidity; it is true rapidity, not pseudorapidity.
Here, $\gamma$ is the Lorentz $\gamma$-factor, $\beta$ is the magnitude of the
three-velocity in units where $c=1$, $\vu$ is the proper velocity and $u$ is its
magnitude.
In exact arithmetic these are interchangeable descriptions. In double
precision they are not, and in the following we outline exactly where the
difference enters.

A 64-bit double stores about $16$ significant decimal digits.\footnote{53-bit significand precision (52 explicit plus one implicit leading bit), an
11-bit exponent field and 1 bit for sign~\cite{IEEE754:2019}.} Near a number of
magnitude $x$ the representable values are spaced by roughly
$\varepsilon|x|$, with $\varepsilon=2^{-52}\simeq2.2\times10^{-16}$; that
spacing is one \emph{unit in the last place}, or \ulp. As a rule of thumb, an
error of $10^{k}\,\ulp$ corresponds to losing roughly $k$ of those 16 decimal
digits.

Converting a velocity into a Lorentz $\gamma$-factor amplifies such errors without
bound. Differentiating $\gamma=(1-\beta^{2})^{-1/2}$,
\begin{equation}
\frac{\mathrm{d}\gamma}{\mathrm{d}\beta}=\beta\left(1-\beta^{2}\right)^{-3/2}
=\gamma^{3}\beta,
\qquad\text{so}\qquad
\frac{\mathrm{d}\gamma}{\gamma}=\gamma^{2}\beta\,\mathrm{d}\beta .
\label{eq:amplify}
\end{equation}
A velocity known only to its last stored digit, $\mathrm{d}\beta\sim
\varepsilon$, therefore fixes $\gamma$ only to a relative error
$\sim\gamma^{2}\varepsilon$. At $\gamma=10^{6}$ the amplification factor
$\gamma^{2}=10^{12}$ leaves only four of the sixteen digits guaranteed to be correct.
In code, the amplification happens in the subtraction $1-\beta^2=1/\gamma^2$, whose
result is far smaller than its operands: a \emph{catastrophic cancellation}.

Pushed far enough, the representation fails outright. Writing
$1-\beta=(1-\beta^{2})/(1+\beta)=1/[\gamma^{2}(1+\beta)]\simeq1/(2\gamma^{2})$,
and noting that the predecessor of unity is separated from it by
$\varepsilon/2\simeq1.1\times10^{-16}$, round-to-nearest produces exactly
$1$ once $1/(2\gamma^{2})$ is below half that spacing, i.e.\ once
\begin{equation}
\gamma \gtrsim \sqrt{2/\varepsilon}\simeq 9.49\times10^{7}
\qquad(\eta\gtrsim 19.06).
\label{eq:saturation}
\end{equation}
In this scalar example, the stored velocity is then exactly $1.0$ and $\gamma=1/\sqrt{0}$ is
infinite. Beyond \eqref{eq:saturation} the velocity carries no information
about the boost magnitude, leaving only information about its sign, and no
algorithm can recover information its input does not contain. All of this is
a property of the representation, present before a single line of boost code
has run.

\subsection*{(b) Algorithmic error}
Given the doubles actually available at a call site, the arithmetic of the boost
adds rounding of its own; unlike (a), this part is under the programmer's control.
Isolating it requires a reference built from the \emph{same rounded inputs} the
algorithm was given, since a comparison with the exact boost reports the sum of (a)
and (b).

\subsection*{(c) Representation loss in the boosted vector}
Rounding the boost input is only half of the representation loss. The same loss
occurs a second time when the \emph{result} of the boost is stored. No algorithm
returning only four Cartesian doubles can recover the information discarded at
that point. The loss is caused by the chosen output representation, not by the
arithmetic that produced it. It is therefore not an alternative to (a) but its
counterpart at the other end of the call.

The conventional four-vector interfaces surveyed in \secref{sec:libs} store
momenta in Cartesian components $(E,\vp)$. What those components can still
resolve after a boost is seen by decomposing the momentum along the boost
direction into light-cone components
\begin{equation}
p^{\pm}=E\pm p_{\parallel},\qquad
m^2=p^+p^- - |\vp_\perp|^2 .
\label{eq:lc-mass}
\end{equation}
A boost of rapidity $\eta$ along $\nhat$ scales $p^\pm\to e^{\pm\eta}p^\pm$
and leaves $\vp_\perp$ unchanged. For large $\gamma$, $e^{\eta}\simeq2\gamma$, so
the ratio $p^+/p^-$ grows as $4\gamma^2$ times its rest-frame value, while their
product, and with it the mass, is untouched. The Cartesian pair
$(E,p_\parallel)$ is only stored to an absolute
precision of about $\varepsilon E\simeq\varepsilon p^+/2$, and the small
component can only be recovered from them as the difference
$p^-=E-p_\parallel$. Its relative error is therefore
\begin{equation}
\frac{\delta p^-}{p^-}\sim\varepsilon\,\frac{p^+}{p^-}
\sim 4\gamma^2\varepsilon\,\frac{p^+_0}{p^-_0},
\label{eq:pminus}
\end{equation}
where the subscript-zero denotes rest-frame values. Through \eqref{eq:lc-mass},
the same relative error enters the invariant mass squared and every other
invariant which depends on $p^-$. The relative error of the mass itself is
smaller by a factor of two when the mass squared remains positive. Minkowski
products and quantities reconstructed after boosting back to a rest frame are
affected in the same way. This is independent of how accurately the boost itself
was evaluated, because it is already there in the correctly rounded result.

The same loss appears when a nearly null system is first constructed from
Cartesian directions. For the pair in \eqref{eq:nearly-null-gamma}, direct
evaluation of $1-\cos\theta$ has a relative error of order
$\varepsilon/\theta^2$. Reconstructing it from
$|\hat{\bm p}_1\mathbin{\times}\hat{\bm p}_2|^2/(1+\cos\theta)$ improves the
generic error to order $\varepsilon/\theta$, because the cross product itself
still subtracts products of rounded components. It does not restore an angle
which has already disappeared from the vectors. If the small transverse
component, the angle or the invariant mass is available independently, it can
instead be retained without either subtraction.

Specialist calculations already retain some of this information. PanScales
stores $m^2$ alongside the Cartesian components of its default momentum type
and can transport differences between nearby directions through
boosts~\cite{Dasgupta:2020fwr}. In the MCFM--Pepper calculation, Bothmann
\emph{et al.} advocate an independently retained virtuality and obtain
collinear information directly from the phase-space variables; Pepper's
general event record remains Cartesian~\cite{Bothmann:2024stable}. Neither
construction retains a general light-cone representation, but both preserve
selected small quantities before Cartesian rounding removes them.

The size is best seen in the system used for the resonance test of
\secref{sec:bench}: a $Z$ at rest, decaying to two muons, boosted along $z$. At
$\gamma=10^{7}$ the pair has $p^+\simeq1.8\times10^{9}\,$GeV and
$p^-\simeq4.6\times10^{-6}\,$GeV, and the individual muons are stored with
energies and longitudinal momenta of order $10^{8}\,$GeV. One \ulp of such an
energy is $\varepsilon E\sim10^{-7}\,$GeV, a few per cent of the whole of $p^-$,
which the mass is then reconstructed from as a difference of those large numbers.
Rounding the boosted energy and longitudinal momentum into doubles --- which even
an exact boost must do --- therefore leaves $p^-$, and with it $m^2=p^+p^-$,
correct to a few per cent only.

Nothing in such an event announces the loss. The components are finite, the
recovered $m^2$ is positive, and a dimuon mass one per cent away from $m_Z$ is an
ordinary entry in a line shape rather than a visibly corrupt number. The failure
only becomes visible to the user when the rounding reaches the size of the invariant 
itself, that is when one \ulp of the energy is as large as $p^-$.

In general, for a relative tolerance $\tol$ on an invariant, an invisible
failure begins once
$\gamma\gtrsim\gamma_{\rm invisible}\simeq\sqrt{\tol/\varepsilon}$. The failure can become visible in the returned value only when the
rounding error reaches the size of the invariant itself, at
$\gamma\gtrsim\gamma_{\rm visible}\simeq\varepsilon^{-1/2}\sim10^{8}$. Between
the two scales,
\begin{equation}
\frac{\gamma_{\rm visible}}{\gamma_{\rm invisible}}
\simeq\frac{1}{\sqrt{\tol}},
\label{eq:invisible-window}
\end{equation}
results are finite and wrong. For $\tol=10^{-3}$ the invisible-failure window
starts at $\gamma\simeq2\times10^{6}$ and spans more than a decade; for
$\tol=10^{-6}$ it spans three.

The practical conclusion is that the dominant error in generic boosts is
usually representation loss, in either the input (a) or the output (c). The
input loss can be removed by supplying the boost through $(\gamma,\vu)$, a
rapidity, or a four-momentum with its mass. The output loss can only be removed
by changing the representation of the boosted vector. Algorithmic error is
smaller in the generic tests, but can become dominant in special kinematics
where a physical component is obtained by subtracting two large terms.

\section{Algorithms}
\label{sec:algos}

In this section we introduce five boost algorithms, labelled A1--A5 with A for
algorithm, for studying numerical behaviour and limitations. The algorithms are briefly listed in \autoref{tab:algorithms}. A stand-alone C++ implementation of the algorithms is included in the accompanying code, see \secref{sec:code} for details. A1, A2 and
A3 are Cartesian boosts that differ in their input: A1 and A2 receive the
three-velocity $\vbeta$, A3 the four-velocity $(\gamma,\vu)$ with $\vu=\gamma\vbeta$.
Of these, A1 and A3 represent prescriptions used in existing software, A2 is a
diagnostic variant, and A4 and A5 are constructive alternatives studied here.
As shown in \secref{sec:error-sources}, $\vbeta$ eventually ceases to resolve the
boost magnitude, whereas $\vu$ continues to grow with $\gamma$. A4 changes the
representation of the boosted vector, and A5 extends that retained
representation to chains of boosts along different axes.

\begin{table}[h!]
\centering
\footnotesize
\begin{tabular}{cll}
\toprule
Label & Input & Numerical prescription \\
\midrule
A1 & $\vbeta$ & Direct boost, with $\gamma$ reconstructed from $1-\beta^2$ \\
A2 & $\vbeta$ & As A1, with compensated evaluation of $1-\beta^2$ \\
A3 & $(\gamma,\vu)$ & Direct proper-velocity boost \\
A4 & $e^{\eta}$, $\nhat$ & Retained light-cone coordinates \\
A5 & chain of $(\gamma,\vu)$ & Composed chain, applied once as A4 \\
\bottomrule
\end{tabular}
\caption{Boost algorithms used in the numerical study. A1, A2 and A3 return
Cartesian components; A4 keeps the boosted vector in coordinates aligned with
the boost axis, and A5 applies this representation after composing a chain. A2
is used to isolate arithmetic error in the component-wise and timing tests.}
\label{tab:algorithms}
\end{table}

\subsection{A1 and A2: velocity input}
\label{sec:family-a}

Decomposing $\vp=p_{\parallel}\nhat+\vp_{\perp}$ into components parallel
and perpendicular to the boost direction $\nhat$, a boost of velocity
$\vbeta=\beta\nhat$ acts as
\begin{equation}
E'=\gamma\left(E+\vbeta\cdot\vp\right),\qquad
p'_{\parallel}=\gamma\left(p_{\parallel}+\beta E\right),\qquad
\vp'_{\perp}=\vp_{\perp}.
\label{eq:a0}
\end{equation}
Writing $\nhat=\vbeta/\beta$ gives
\begin{equation}
\vp'=\vp+\vbeta\left[\frac{\gamma-1}{\beta^{2}}\left(\vbeta\cdot\vp\right)+\gamma E\right] =\vp+\left[\frac{\gamma^{2}}{\gamma+1}\,\vbeta\cdot\vp
      +\gamma E\right]\vbeta.
\label{eq:a1}
\end{equation}
The first grouping, $(\gamma-1)/\beta^2$, used by \root, \clhep and \thepeg (see \secref{sec:libs}), is
algebraically equivalent to the second, but is indeterminate at $\beta=0$ unless
guarded. A1 reconstructs $\gamma=1/\sqrt{\alpha}$ with the textbook
$\alpha=1-(\beta_x^2+\beta_y^2+\beta_z^2)$ and applies \eqref{eq:a1}. As discussed in
\secref{sec:error-sources}, the absolute rounding error of the squared norm becomes a
relative error $\delta\alpha/\alpha=O(\gamma^{2}\varepsilon)$. A2 evaluates the same
$\alpha$ with error-free transformations of the products and sums~\cite{Dekker:1971,Knuth:1997},
which retain their rounding residuals, and otherwise applies the same update. A2
thus removes the algorithmic error of the norm, but cannot recover information
already lost when $\vbeta$ was rounded; it is included to show that removing
algorithmic error does not help once the error is representation loss.

\subsection{A3: four-velocity input}

A3 receives $(\gamma,\vu)$, where
$\gamma^2-u^2=1$.  These four stored numbers contain only three independent
physical quantities, but avoid encoding a large boost in the small
difference between $\beta$ and one.
Using $\vu=\gamma\vbeta$ and
$\gamma-1=u^2/(\gamma+1)$, the direct A3 update is
\begin{equation}
E'=\gamma E+\vu\cdot\vp,\qquad
\vp'=\vp+\vu\left[\frac{\vu\cdot\vp}{\gamma+1}+E\right].
\label{eq:a3}
\end{equation}
Its only division is by $\gamma+1\geq2$, and it reconstructs neither a
velocity norm nor a Lorentz factor.  The relation
$\gamma=\sqrt{1+u^2}$ is widely used in relativistic particle-update
algorithms (``particle pushers'')~\cite{Vay:2008,Schmitz:2026}. Several library interfaces instead accept a
Lorentz factor together with the velocity and apply \eqref{eq:a1}; this avoids the
reconstruction of $\gamma$ in the same way. It therefore avoids the input loss,
although its Cartesian result remains exposed to the output loss. It can also
suffer algorithmic cancellation when a small physical component is formed as
the difference of the two terms in \eqref{eq:a3}, as the antiparallel benchmark
below demonstrates.

\subsection{A4: retained coordinates}
\label{sec:retained}

For $u=|\vu|>0$, define $\nhat=\vu/u$,
$p_{\parallel}=\nhat\cdot\vp$ and $p^\pm=E\pm p_{\parallel}$.  The two
light-cone components transform independently.  Writing
$\lambda=e^\eta=\gamma+u$ and using
$(\gamma-u)(\gamma+u)=1$, A4 evaluates
\begin{equation}
p'^+=\lambda p^+,\qquad p'^-=\frac{p^-}{\lambda},\qquad
\vp'_{\perp}=\vp_{\perp}.
\label{eq:lightcone}
\end{equation}
The difference to a Cartesian boost is not the formula but what is kept
afterwards. A4 converts the vector once to $(p^+,p^-,\vp_\perp)$, in a frame where
this is benign, and keeps it in that form while it is boosted. Invariants are
evaluated from \eqref{eq:lc-mass}, which subtracts no large components. A Cartesian
vector is recovered as $E'=(p'^++p'^-)/2$ and $p'_{\parallel}=(p'^+-p'^-)/2$ only on
request. This conversion loses invariants when the two light-cone components
can no longer be resolved simultaneously as Cartesian doubles. It remains
benign when the resulting Cartesian components are themselves small, as in the
antiparallel benchmark below. The construction solves boosts along a single
axis. A chain of boosts along different axes, or a rotation, cannot be expressed
as a rescaling, and must leave the light-cone representation.

\subsection{A5: chains of boosts along different axes}
\label{sec:chains}

A boost along a different axis cannot be written as a rescaling of light-cone
components along the first one, so a chain of non-collinear boosts must leave the
retained representation between steps, and every Cartesian step is again subject to
the output loss of \secref{sec:error-sources}(c). A5 avoids this by composing the
chain before it is applied.

A Lorentz transformation can always be written as a rotation followed by a
boost, $T=B(\vu)\,R$. Two boosts compose into a boost and a Wigner
rotation~\cite{Berry:2021byt},
\begin{equation}
B(\bm{a})\,B(\bm{b})=B(\bm{c})\,W,\qquad
\gamma_c=\gamma_a\gamma_b+\bm{a}\cdot\bm{b},\qquad
\bm{c}=\bm{b}+\bm{a}\left[\frac{\bm{a}\cdot\bm{b}}{\gamma_a+1}+\gamma_b\right],
\label{eq:compose}
\end{equation}
where $\bm{c}$ is \eqref{eq:a3} applied to the four-velocity $(\gamma_b,\bm{b})$,
and $W$ is a rotation about $\bm{b}\times\bm{a}$ by the angle $\omega$ with
\begin{equation}
\sin\omega=\frac{K\,|\bm{a}\times\bm{b}|}{D},\qquad
1+\cos\omega=\frac{K^2}{D},\qquad
K=1+\gamma_a+\gamma_b+\gamma_c,\quad D=(1+\gamma_a)(1+\gamma_b)(1+\gamma_c).
\label{eq:wigner}
\end{equation}
Cross products are also used to retain small angular differences in numerical
parton-shower constructions~\cite{Dasgupta:2020fwr,Herren:2022jej}. They do not
by themselves recover a difference already lost from rounded Cartesian
vectors, as the nearly null benchmark below shows. In A5, the factor
$1-\cos\omega$ in the rotation matrix is evaluated as
$\sin^2\omega/(1+\cos\omega)$ when $\cos\omega\geq0$, avoiding cancellation
near $\omega=0$. For $\cos\omega<0$, A5 uses $1-\cos\omega$ directly, avoiding
the singular denominator of the rearranged expression at $\omega=\pi$. The identity
\begin{equation}
B(\bm{c}_2)R_2\,B(\bm{c}_1)R_1
=B(\bm{c}_2)B(R_2\bm{c}_1)R_2R_1
\end{equation}
then reduces a chain of any length to a single $T=B(\vu)R$ by repeated use of
\eqref{eq:compose}. A5 composes the chain once, applies $R$ to the vectors in
their original frame, where rotations are benign, and performs the single remaining
boost with A4 along $\hat{\vu}$. The inverse is the A4 inverse boost followed
by $R^{\mathsf T}$. The vectors never pass through a Cartesian representation
in a highly boosted frame.

Composition has its own limit. When a later boost nearly undoes an earlier
one, $\bm{a}\cdot\bm{b}\simeq-\gamma_a\gamma_b$, and \eqref{eq:compose} forms a
small $\gamma_c$ from large terms. This is algorithmic cancellation in the
composition of the transformation, rather than representation loss in a
boosted vector. A5 also
requires the whole chain to be known while the vectors are still in a benign
frame, and invariants between vectors with different chains need a common frame,
in which the cancellation can return.

\section{Library implementations}
\label{sec:libs}

The failure modes of \secref{sec:error-sources} are properties of interfaces and
representations, so whether they matter in practice depends on how existing code
specifies its boosts, how it calls them and what it stores. We surveyed this from
the source code of the collider generators, analysis tools, two air-shower codes
and a hadronic cascade code listed in \autoref{tab:libs}. \autoref{tab:calls}, in turn, lists how the
packages call them. The tables are restricted to established packages exposing
general-purpose four-vector or boost interfaces; the more specialist PanScales and
MCFM--Pepper constructions are not included in the survey.

The survey was carried out by an AI agent (Claude Opus 5) from a fixed prompt,
which pins the source versions, defines the classification and requires a file
and line reference for every statement; prompt and dated output are part of the
code release (\secref{sec:code}). The agent's survey results were checked by the
author as follows: the identified input and formula family were manually
cross-checked at the indicated place in the code, as were the indicated guards.
Call sites were spot-checked, but the total count of surveyed call sites was not
independently confirmed and should therefore be taken as approximate. An interested reader can use the included prompt to extend the survey to a code of their choice.

The discussion below relates their interfaces and
representations to the failure modes established in \secref{sec:error-sources}, and thus to the benchmark algorithms A1--A5.

\begin{table}[t]
\centering
\footnotesize
\resizebox{\textwidth}{!}{%
\begin{tabular}{llll}
\toprule
Library & Routine & Input & Formula family\\
\midrule
\pythia 8.317~\cite{Bierlich:2022pfr} & \texttt{Vec4::bst($\beta_x,\beta_y,\beta_z$)} & $\vbeta$ & A1, $g=\gamma^2/(\gamma+1)$, guard\footnotemark\\
\pythia 8.317 & \texttt{Vec4::bst(pIn, mIn)} & $(P, M)$ & A1 with $\gamma=E/M$ supplied, $\vbeta=\vp/E$\\
\root 6.37/01~\cite{Brun:1997pa} / \clhep 2.4.2.0~\cite{Lonnblad:1994np} & \texttt{TLorentzVector::Boost} & $\vbeta$ & A1, $g=(\gamma-1)/\beta^2$\\
SMASH 2.2.1+~\cite{SMASH:2016zqf} & \texttt{FourVector::lorentz\_boost} & $\vbeta$ & energy-first form of A1, guard\footnotemark[\value{footnote}]\\
\vect 1.7.0+~\cite{Chopra:2025faa} & \texttt{boost\_p4} & $P$ & matrix, $\gamma=E/M$, $M^2=E^2-|\vp|^2$\\
\rivet 4.1.3~\cite{Bierlich:2024vqo} & \texttt{LorentzTransform::setBetaVec} & $\vbeta$ & matrix, conjugated by rotations\\
\rivet 4.1.3 & \texttt{LorentzTransform::setGammaVec} & $\gamma\nhat$ & same, $\beta=\sqrt{1-1/\gamma^{2}}$\\
\fastjet 3.5.1~\cite{Cacciari:2011ma} & \texttt{PseudoJet::boost} & $P$ & A3 (energy first), $M$ recomputed\\
\thepeg 2.3.0~\cite{Lonnblad:2006pt} / \herwig 7.3.0~\cite{Bewick:2023tfi} & \texttt{LorentzVector::boost(b,\,gamma)} & $(\vbeta{=}\vp/E,\gamma{=}E/M)$ & A1 with $\gamma$ supplied, $g=(\gamma-1)/\beta^2$\\
\sherpa 3.1.x~\cite{Sherpa:2024mfk} & \texttt{Poincare::BoostBack}, \texttt{rsq} $=M$ & $(P, M)$ & A3 (energy first)\\
\sherpa 3.1.x & same, default \texttt{rsq} & $P$ & A3 (energy first), $M$ recomputed\\
\addlinespace
\corsika 8~\cite{Engel:2018akg} & \texttt{COMBoost} & $(P,M_{\rm target})$, $P_1{+}P_2$ or $(|\vp|,M)$ & rotation, A3 on $(E,p_z)$\\
EPOS 4.0.0~\cite{Werner:2023zvo} & \texttt{utlob2} (\texttt{utlob3}, \texttt{utlob4}) & $(P,M)$ & A3, $E$ reset from input invariant\\
EPOS 4.0.0 & boost in \texttt{hdecay} & $(P,M)$ & A1 with $\gamma=E/M$ supplied, single precision\\
EPOS 4.0.0 & \texttt{utlob5} & rapidity & rapidity shift, single precision\\
\bottomrule
\end{tabular}
}
\caption{Surveyed boost routines, the input they receive and their arithmetic, in
the notation of \autoref{tab:algorithms}; $g$ is the coefficient of
$\vbeta(\vbeta\cdot\vp)$ in the A1 update. \root, \clhep, SMASH and \vect were
surveyed at commits after the stated releases, \sherpa at a commit of the 3.1.x
branch. \root's \texttt{TLorentzVector::Boost}, \clhep's
\texttt{HepLorentzVector::boost} and \thepeg's \texttt{LorentzVector::boost} without a
supplied $\gamma$ consist of the same statements. \sherpa's
\texttt{Poincare} performs the same operations as \fastjet apart from the mass.
The \vect entry assumes a booster stored
with an energy coordinate.}
\label{tab:libs}
\end{table}
\footnotetext{\pythia returns the vector unchanged when $\beta^{2}\geq1$; SMASH sets $\gamma=0$ when $\beta^2\geq1$.}

\begin{table}[t]
\centering
\footnotesize
\begin{tabular}{llll}
\toprule
Package & Call sites surveyed & $\gamma$ or $M$ supplied & $M$ or $s$ from $E^2-|\vp|^2$\\
\midrule
\pythia 8.317 & 192 \texttt{bst}/\texttt{bstback} & 29 & 10 of the 29\\
\herwig 7.3.0 & 291 \thepeg boosts & 30 & at some $\vbeta$ formulas\\
\rivet 4.1.3 & 654 transform factories & 3 & beam masses, at those 3\\
\sherpa 3.1.x & 182 \texttt{Poincare($P$)} & 0 & all 182\\
SMASH & 26 & 0 & --\\
\fastjet 3.5.1 & 6 (in \texttt{tools}) & 0 & all 6\\
\corsika 8 & 17 \texttt{COMBoost} & 3 & 14\\
EPOS 4.0.0 & 53 \texttt{utlob} calls & $M$ always & 25 at the call site, 2 more indirectly\\
\bottomrule
\end{tabular}
\caption{Call patterns of the surveyed boosts in the packages' own code, counted
in the survey. ``$\gamma$ or $M$ supplied'' counts sites at which a Lorentz factor or
the mass of the boosted system is passed rather than reconstructed. The last column
counts sites at which that mass, or $s$, is formed from Cartesian components.}
\label{tab:calls}
\end{table}

\subsection{Interfaces and their use}

Every surveyed routine returns Cartesian components, so all of them are exposed to
the output loss of \secref{sec:error-sources}(c), but they differ in whether they also
suffer the input loss (a). \root,
\clhep and SMASH accept only a velocity, and any code built on them reconstructs
$\gamma$ from $1-\beta^2$ and inherits the velocity saturation, whatever
information the caller had.

Other packages, among them \pythia, \thepeg and
\sherpa, also accept a Lorentz factor or the mass of the boosted system. \thepeg
documents its optional $\gamma$ as being ``for higher numerical accuracy''. The call
patterns of \autoref{tab:calls} show that these interfaces are the exception in
practice. \pythia passes
$\gamma$ or a mass at 29 of its 192 boost calls. In \herwig, only 24 of the 247
calls that boost by $\vbeta=\vp/E$ also pass $\gamma=E/M$ from the stored mass.
The other 223 let $\gamma$ be reconstructed with A1 arithmetic. The remaining
44 \herwig calls construct the velocity in other ways, and six of these supply
a Lorentz factor, accounting for the totals of 291 calls and 30 supplied
Lorentz factors in \autoref{tab:calls}.

\sherpa's interface accepts the mass, but none of its 182 calls uses it, and \rivet
passes a Lorentz factor only when constructing the beam centre-of-mass frame. The
interfaces in \pythia, \thepeg and \sherpa that avoid the input loss therefore
describe a minority of the boosts actually performed. None of the surveyed routines evaluates $1-\beta^2$ with compensated
arithmetic, which, as A2 shows, would not change this.

\subsection{Where the mass comes from}

An interface that receives only the four-momentum of the boosted system has to
recover its mass as $\sqrt{E^2-|\vp|^2}$ from Cartesian components, which is the
output loss applied to the input of the boost. This happens at every call of
\fastjet's boost, of \vect's \texttt{boost\_p4} for an energy-coordinate booster and,
given the call patterns, of \sherpa's \texttt{Poincare}. \sherpa takes the absolute
value of the recomputed $m^2$, so that even a mass lost entirely to rounding yields
a finite boost. This also happens in some code
that nominally supplies the mass, but really just hides the problem. For example, at 10 of \pythia's 29 calls, the mass or Lorentz factor passed was itself computed from Cartesian components. Supplying the mass only helps if it is taken from a stored value, not if it is calculated in the same way as the lossy routine would do it.

The survey also finds the reverse: stored masses that are overwritten after a
transformation. Boosting a \root GenVector held in a mass-based coordinate system
converts it to Cartesian components, boosts it, and recomputes the mass from the
result, and \thepeg's rotation applied to a five-vector does the same. Storing the
mass alongside the momentum would protect single-particle invariants, as the
air-shower code discussed below shows, but these routines replace it with the lossy
value exactly when the boost is large enough for the difference to matter.

\subsection{What the guards do}

The velocity guards in the surveyed code test $\beta^2\geq1$, and therefore
act only at the visible boundary of the invisible-failure window of
\eqref{eq:invisible-window}, if at all. What they do when triggered decides
whether the failure stays visible.
\clhep's boost classes and \thepeg's rotation matrices throw an exception, which is
the behaviour a user needs. \root's \texttt{TLorentzVector::Boost}, \clhep's
\texttt{HepLorentzVector::boost}, \thepeg's \texttt{LorentzVector::boost} and \vect
have no test and return non-finite components once the velocity saturates, which is
at least noticeable. Several
routines, however, replace the failure by a finite result. \pythia's
\texttt{Vec4::bst} returns the vector unboosted, so that invariants remain exactly
right while the kinematics are wrong; its
matrix boost instead caps $\gamma$ at $10^{10}$. SMASH sets $\gamma=0$. \root's
GenVector boosts report the condition through an exception mechanism that is
switched off by default, after which they return a zero vector or leave the boost
matrix unset.

\rivet's guard is of a different kind. It returns the identity whenever
$|\vbeta|^2$ or $|\gamma^2-1|$ is below $10^{-8}$, that is, for every boost with
$\beta\lesssim10^{-4}$. This is not a floating-point limit, but rather a minimal threshold under which a boost would be considered inconsequential. However, a boost of that size
applied to an already highly boosted system can change its kinematics substantially. The
guard can therefore replace a physically relevant transformation by the identity.

\subsection{Air-shower codes}

\corsika~8 performs its Lorentz transformations with \texttt{COMBoost}, which rotates
the boost direction onto the $z$ axis and boosts $(E,p_z)$: the A3 form with
Cartesian output. The boosts are therefore subject to the output loss, and the code
acknowledges it. The particle stack stores the kinetic energy and
the direction of a particle and takes its mass from the particle table, and at every
site where secondaries are transformed back from the centre-of-mass frame, the
boosted energy is discarded and recomputed from the momentum and the table mass. One
of these sites explains that the boosted energy ``can be smaller than the momentum
due to limited numerical accuracy''. By returning every secondary to its mass shell,
the stack protects single-particle propagation. What it cannot protect are quantities built from several
particles, and the Cartesian inputs of the boost itself, where \texttt{COMBoost}
forms the projectile mass or the pair invariant $s$ from $E^2-|\vp|^2$ at 14 of its
17 constructions.

In the surveyed EPOS~4.0.0 release, the particle record,
$\texttt{pptl}=(p_x,p_y,p_z,E,m)$, is stored in single precision. The
EPOS-LHC and EPOS-LHC-R particle records shipped with the surveyed \corsika~8
revision are likewise single precision. The other surveyed interaction-model
particle records in \corsika~8 use double precision. With
$\varepsilon_{32}\simeq1.2\times10^{-7}$, the whole analysis of
\secref{sec:error-sources} moves down. The invisible-failure window of
\eqref{eq:invisible-window} opens at
$\gamma\simeq\sqrt{\tol/\varepsilon_{32}}\simeq90$ for $\tol=10^{-3}$, far below
the Lorentz factors reached in air showers. Decay products are boosted and
stored in single precision, and the transformation of the final event into the
target frame, although it is a shift of rapidity computed with the stored mass,
converts straight back to single-precision Cartesian components and so gives up the
advantage of the rapidity representation. The double-precision boost \texttt{utlob2} shows how a correction
meant to enforce the mass shell becomes a failure path of its own. After the boost, it
resets the energy to $\sqrt{m_{\rm in}^2+|\vp'|^2}$, with $m_{\rm in}^2$ computed from
the Cartesian components of the input. When the input vector is itself highly
boosted, $m_{\rm in}^2$ is precisely the quantity lost to cancellation, and the reset
transfers that error to the result. To further obscure the failure path, if the argument of the square root is not positive, the energy is set to zero, with a message only at a debugging print level.
The single-precision wrappers of this routine additionally round every component to
about seven significant decimal digits before the double-precision arithmetic
starts, and at about half of all calls the mass passed in is computed from Cartesian
components. EPOS is compiled to trap invalid floating-point operations, so failures
that produce NaN values end the run, while none of the invisible paths above does.

The survey therefore exposes a consistent pattern. Velocity-only entry points
remain common, while interfaces capable of receiving a Lorentz factor or mass
are often used without it or are supplied a value reconstructed from Cartesian
components. Every surveyed routine ultimately returns Cartesian components.
Keeping a separately stored mass can preserve the mass shell of an individual
particle, as in the \corsika~8 stack, but does not preserve general multiparticle
invariants or observables reconstructed in another frame. The benchmarks below
therefore test the boost input and the representation of its output separately.

\section{Benchmarks}
\label{sec:bench}

Where a numerical reference is required, the generated rest-frame vectors and
the requested Lorentz factor are promoted to double-double precision
($\sim32$ significant digits), boosted with the A3 formula, and kept as
Cartesian double-double components throughout the
calculation~\cite{Dekker:1971,Knuth:1997,HidaLiBailey:2001}. Such curves are labelled ``reference'' in the following.

The standard resonance and chain scans use $\gamma=10^t$ for
$t=0,0.25,\ldots,10$, with additional points at the physical scales of
\secref{sec:physical-scales} and around the velocity saturation
\eqref{eq:saturation}. The random component test extends to $10^{12}$, the
extreme resonance test to $10^{17}$, and the analytic accuracy and scaled-chain
tests to $10^{50}$.

\begin{description}
\setlength{\itemsep}{2pt}
\item[Component-wise accuracy.] For each $\gamma$, 256 reproducibly generated
timelike, lightlike and spacelike vectors are boosted in random directions.
The largest component error is measured in \ulp of the largest reference
component. The \emph{total} error uses the intended double-double boost, while
the \emph{algorithmic} error uses a double-double reference constructed from
the rounded inputs actually received by each algorithm. Their separation is
the input representation loss of \secref{sec:error-sources}(a). A second test
uses a massless vector with $\vp=-E\nhat$, exactly antiparallel to the boost.
It represents a backward particle which is red-shifted to
$E'=e^{-\eta}E$, and directly probes whether the small light-cone component
survives the boost.

A third test constructs the pair of \eqref{eq:nearly-null-gamma} with unit
particle energies and $\theta=2\arcsin(1/\gamma_{12})$. Its exact invariant is
$M^2=4/\gamma_{12}^2$. We compare direct Cartesian reconstruction, the stable
cross-product identity applied to generic rounded directions, a construction
which retains the small transverse angular component explicitly, and an
independently retained $M^2$.

\item[Resonance.] A $Z$ with a Breit--Wigner mass distribution
($m_Z=91.1880\,$GeV, $\Gamma_Z=2.4955\,$GeV, $60<m<120\,$GeV) decays
isotropically to two massless muons. The two momenta are boosted along $z$ and
their invariant mass is reconstructed in the boosted frame. The plotted
quantity is the event-by-event relative RMS
\begin{equation}
 R_m=\left[\frac{1}{N_{\rm fin}}\sum_{i=1}^{N_{\rm fin}}
 \left(\frac{m_i^{\rm rec}-m_i^{\rm gen}}{m_i^{\rm gen}}\right)^2\right]^{1/2},
 \label{eq:rms-mass}
\end{equation}
over events with a finite reconstructed mass. A separate sample of
$\Upsilon(1S)\to\mu^+\mu^-$, with $\Gamma_\Upsilon=54.02\,$keV, shows how the
same error distorts a narrow line shape at $\gamma=5\times10^5$.
The scan is extended through $\gamma=10^{17}$ and includes the three
ultra-heavy-parent scales of \eqref{eq:heavy-parent-gammas}.

\item[Boost chains.] A $100\,$MeV virtual photon decays to an effectively
massless $e^+e^-$ pair with a $1+\cos^2\theta^*$ distribution. The pair is
boosted through a chain and back through its inverse, after which
$\cos\theta^*$ is reconstructed. The first two chains contain perpendicular
boosts in the two possible orders: $\gamma_x=10$ then $\gamma_z$, and
$\gamma_z$ then $\gamma_x=10$. The fixed transverse boost makes the combined
Lorentz factor $10\gamma_z$ and exposes the ordering of the retained-coordinate
conversion. The third chain applies
$\gamma_z$, a transverse kick $u_x=1/\gamma_z$, and the inverse $z$ boost. Its
net Lorentz factor is about $1.5$, although its intermediate frame is highly
boosted. A fourth chain omits the inverse $z$ boost:
$B_x(u_x=1/\gamma_z)B_z(\gamma_z)$. Its longitudinal boost remains large, its
transverse effect stays finite because $u_x\gamma_z=1$, and its Wigner angle
decreases as $1/\gamma_z$. A5 composes each chain once before it is applied.
\end{description}

The resonance and chain tests use $3\times10^5$ events per point. They also
record non-finite outcomes. The chain test records the RMS error in
$\cos\theta^*$ and, for the nearly cancelling chain, the lab-frame energy before
the inverse transformation; the latter exposes errors which a round trip can
cancel. For the scaled chain, the first particle's transverse momentum after
the forward transformation is also compared with the analytic result
\begin{equation}
 p_x'=\gamma_xp_x+E+\beta_zp_z,\qquad
 \gamma_x=\sqrt{1+\gamma_z^{-2}},
 \label{eq:scaled-chain-px}
\end{equation}
which does not construct the large intermediate energy. The timing test reports
the median of seven passes at $\gamma=10^6$.

\begin{figure}[h!]
\centering
\includegraphics[width=0.48\textwidth]{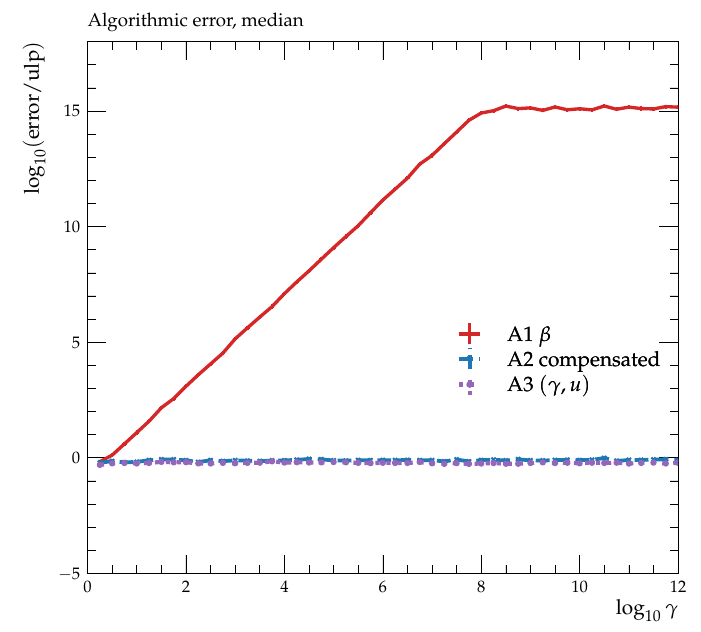}\hfill
\includegraphics[width=0.48\textwidth]{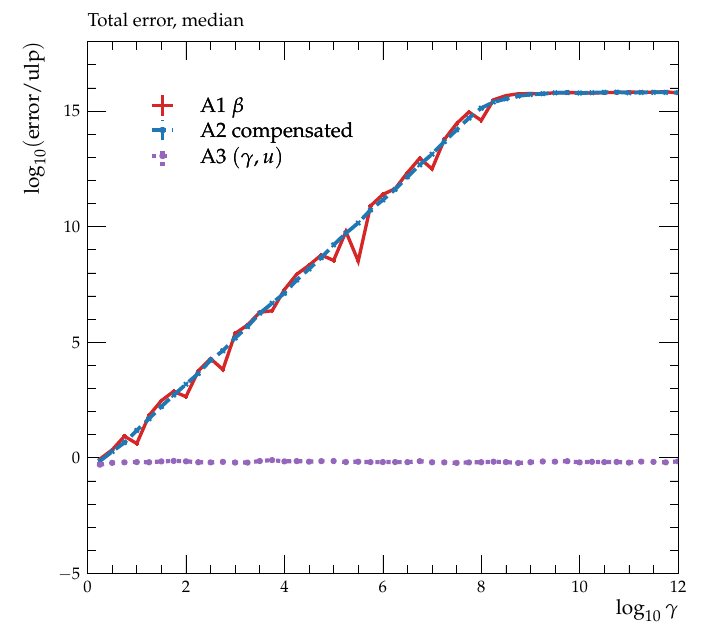}
\caption{Component-wise accuracy. Left: median arithmetic error relative to a
double-double calculation using the same rounded input. Right: median total
error relative to the intended boost. Both use the largest component error in
\ulp of the largest reference component. Non-finite points are omitted.}
\label{fig:accuracy}
\end{figure}

\section{Results}
\label{sec:results}

\subsection{Component-wise accuracy}

The two panels of \figref{fig:accuracy} separate algorithmic from representation error. Compensating the evaluation of $1-\beta^2$ reduces A2's
algorithmic error below one \ulp, but does not improve its total error over A1, since the rounded velocity has already lost the boost magnitude. A3 avoids that input
loss and remains at about one \ulp through $\gamma=10^{12}$. The clear conclusion from this test, is that compensated evaluation is not a viable strategy.

The rise in the right panel of \figref{fig:accuracy} checks the estimate of
\secref{sec:error-sources} directly. Since one \ulp corresponds to a relative
scale of order $\varepsilon$, the predicted $\gamma^2\varepsilon$ error is of
order $\gamma^2\,\ulp$; between $\gamma=10$ and $10^6$, A1 and A2 indeed rise
by about ten decades.

\begin{figure}[t]
\centering
\includegraphics[width=0.48\textwidth]{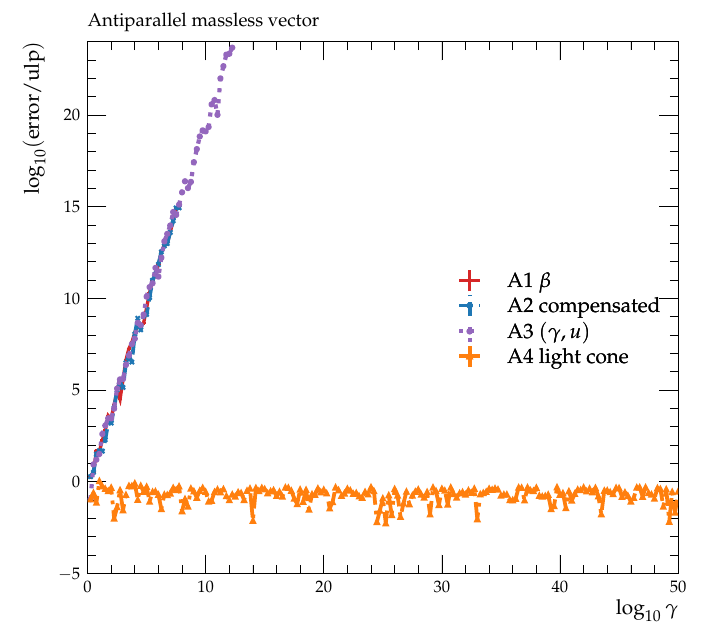}\hfill
\includegraphics[width=0.48\textwidth]{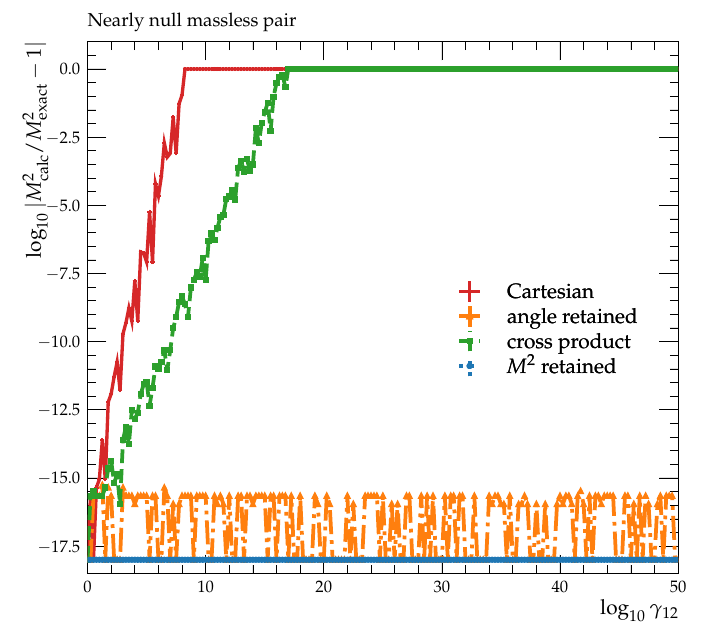}
\caption{Special configurations. Left: an antiparallel massless vector exposes
the red-shifted component lost by Cartesian A3 but retained by A4. Right:
relative error in $M^2=4/\gamma_{12}^2$ for two nearly collinear massless
vectors. Values below
$10^{-18}$ are drawn at $10^{-18}$.}
\label{fig:special-accuracy}
\end{figure}

The left panel of \figref{fig:special-accuracy} is the antiparallel boost,
which explicitly tries to break A3. For an exactly antiparallel massless
vector, the direct Cartesian formulas form the red-shifted energy from the
difference of two terms of order $\gamma E$. A3 therefore loses digits even
though it received an accurate boost, reaching about $10^{11}\,\ulp$ at
$\gamma=10^6$. A4 evaluates the same transformation as a light-cone rescaling
and remains at about one \ulp.

The nearly null test in the right panel shows where the small invariant
is lost before any boost is performed. Direct Cartesian reconstruction exceeds
one per cent error near $\gamma_{12}=2\times10^7$ and loses the invariant near
$10^8$. The cross-product construction with generic rounded directions delays
the same boundary to $\gamma_{12}\sim10^{15}$--$10^{16}$, but does not remove
it. When the small transverse angular component is represented explicitly, the result remains at rounding precision through $\gamma_{12}=10^{50}$, i.e. similar to the reference curve where $M^2$ is retained independently.

\subsection{Resonance}

\begin{figure}[t]
\centering
\includegraphics[width=0.48\textwidth]{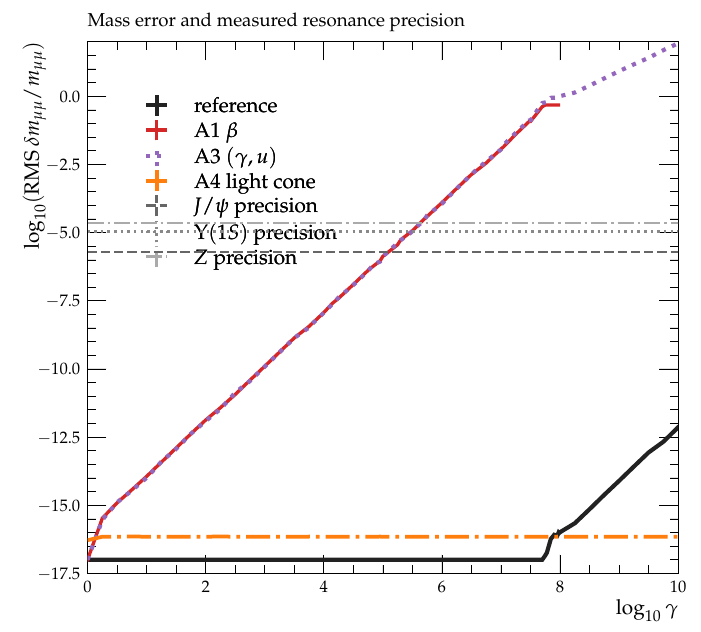}\hfill
\includegraphics[width=0.48\textwidth]{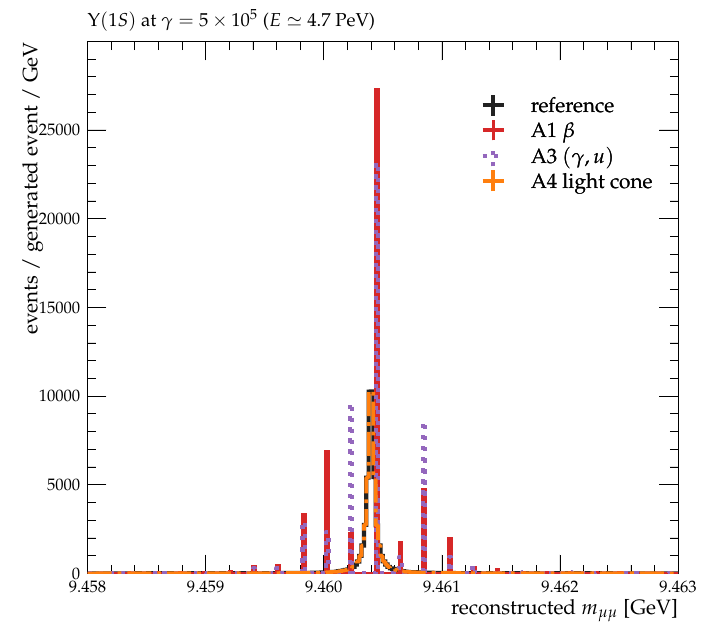}
\caption{Resonance test, with $3\times10^5$ events per curve. Left: relative
event-by-event RMS mass error $R_m$ for the boosted $Z$ sample, evaluated over
finite reconstructed masses. The horizontal lines show the current fractional
mass uncertainties of \autoref{tab:resonance-precision}. The late rise of the
double-double Cartesian reference is the same representation loss shifted to
higher $\gamma$. Right: reconstructed $\Upsilon(1S)$ line shape at
$\gamma=5\times10^5$ ($E\simeq4.7\,$PeV), normalised to the generated event
count. A1 and A3 form a discrete mass comb, while the reference and A4 retain
the generated line shape.}
\label{fig:resonance}
\end{figure}

The left panel of \figref{fig:resonance} shows $R_m$ from
\eqref{eq:rms-mass}. A1 and A3 overlap while they return finite masses. The
dominant error is the Cartesian output, not the boost input. It grows as
$\gamma^2$ and reaches the current fractional mass uncertainties of the
$J/\psi$, $\Upsilon(1S)$ and $Z$ at Lorentz factors of order
$10^5$, $3\times10^5$ and $4\times10^5$, respectively. The horizontal lines are
scale markers from \autoref{tab:resonance-precision}.

The double-double reference begins to rise between $\gamma=6\times10^7$ and
$10^8$, and then develops the same $\gamma^2$ growth. More precision delays the
loss but leaves the Cartesian cancellation in place. A4 instead stays at the
precision of the generated rest-frame vectors throughout the scan.

The right panel shows the distortion of the $\Upsilon(1S)$ line shape. $R_m$
for $\Upsilon(1S)$ crosses the uncertainty on the fitted mass near $3\times10^5$, and develops into the displayed comb at the displayed value
$\gamma=5\times10^5$.

Finite output is a much weaker condition. At $\gamma=10^7$, A1 and A3 still
return a finite mass for every event, but their RMS mass errors are already
about $1\,$GeV. More than one per cent of A3 events become non-finite at
$\gamma=5.6\times10^7$; A1 crosses that fraction only when its velocity
saturates near $9.5\times10^7$. A4 returns finite masses at rest-frame precision
throughout the scan.

\begin{figure}[t]
\centering
\includegraphics[width=0.48\textwidth]{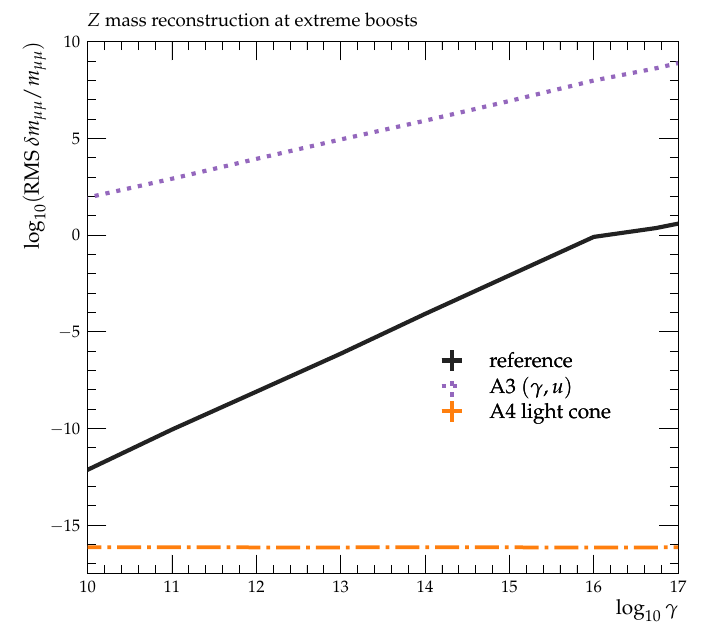}
\caption{The $Z\to\mu^+\mu^-$ mass reconstruction of
\figref{fig:resonance}, extended to the hierarchy of ultra-heavy parent
decays. A4 retains light-cone coordinates.
No A1 curve is visible because it returns no finite reconstructed masses in
the displayed range.
The RMS is evaluated over finite reconstructed masses; non-finite events are
excluded from the RMS rather than assigned an error.}
\label{fig:resonance-extreme}
\end{figure}

The extension in \figref{fig:resonance-extreme} follows the same $Z$ test to
the largest scale in \eqref{eq:heavy-parent-gammas}. The double-double
Cartesian reference eventually suffers the same output loss: its relative RMS
mass error is of order one at $\gamma\sim10^{16}$. A4 retains the two
light-cone components and stays at the precision of the generated rest-frame
sample through $\gamma=10^{17}$. The comparison is therefore between
representations, not between ordinary and insufficiently careful evaluations
of the same Cartesian formula.

\subsection{Boosts along different axes}
\label{sec:results-chains}

\begin{figure}[htbp]
\centering
\includegraphics[width=0.48\textwidth]{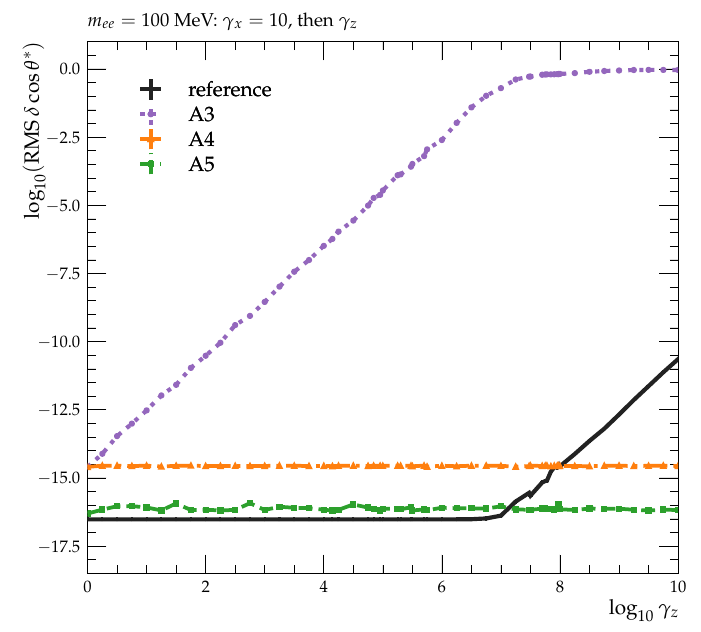}\hfill
\includegraphics[width=0.48\textwidth]{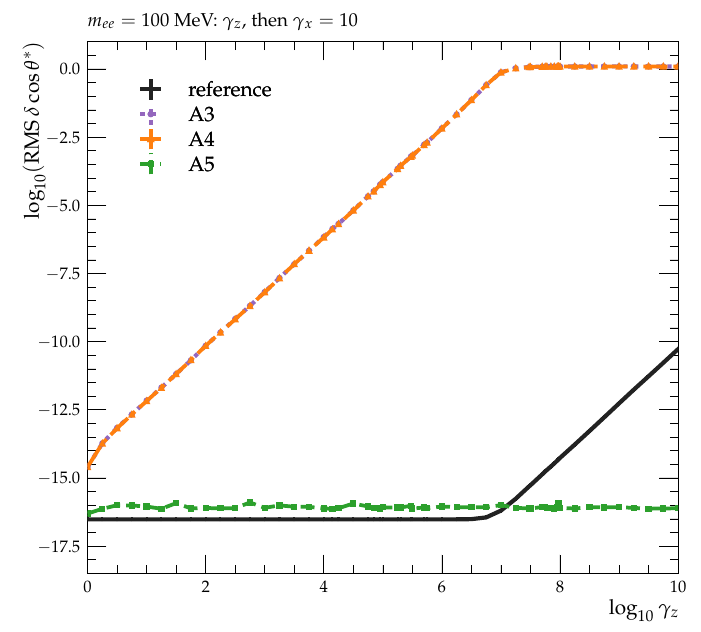}
\caption{Boost-chain test for a $100\,$MeV virtual photon, using
$3\times10^5$ decays per point. The plotted angle error is measured after the
chain and its inverse. The panels show perpendicular boosts with fixed
$\gamma_x=10$ in the two possible orders; this gives total Lorentz factor
$10\gamma_z$ and makes the cost of leaving retained $z$ coordinates visible.}
\label{fig:chain-order}
\end{figure}

The two panels of \figref{fig:chain-order} differ only in the order of the two
perpendicular boosts. Their combined Lorentz factor is
$\gamma_x\gamma_z=10\gamma_z$. A3 therefore loses the rest-frame angle in both
cases, with an RMS error of $2$--$7\times10^{-3}$ at $\gamma_z=10^6$ and of
order one by $\gamma_z\sim10^8$. A4 is accurate when the large $z$ boost is
last. It performs the transverse boost in the original frame and retains
light-cone coordinates afterwards. When the large boost is first, the later
transverse boost forces A4 back through Cartesian components and it follows A3.
A5 composes the non-cancelling perpendicular boosts and remains at rest-frame
precision in either case.

\clearpage

\begin{figure}[htbp]
\centering
\includegraphics[width=0.48\textwidth]{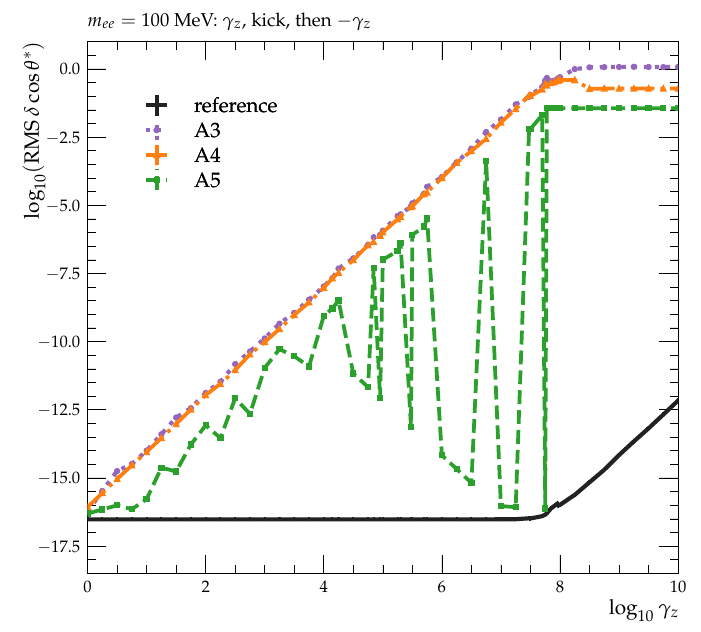}\hfill
\includegraphics[width=0.48\textwidth]{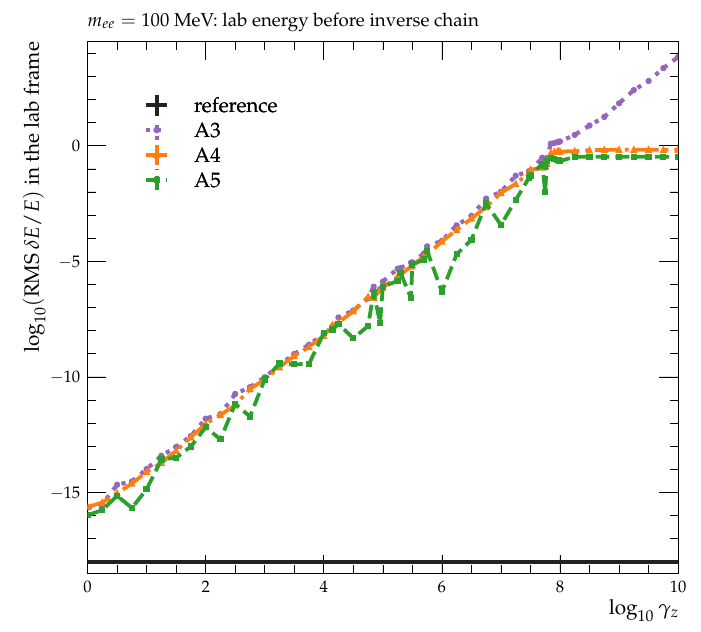}
\caption{Near-cancelling chain for the same $100\,$MeV virtual-photon decays
as \figref{fig:chain-order}: a boost $\gamma_z$, a transverse kick
$u_x=1/\gamma_z$, and the inverse $z$ boost. Left: the RMS rest-frame angular
error after the complete round trip. The jagged A5 curve is caused by
cancellation while composing the transformation, not by event statistics.
Right: relative lab-frame energy error after the forward chain and before its
inverse; it exposes errors which cancel in the round trip. Values below
$10^{-18}$ are drawn at $10^{-18}$.}
\label{fig:chain-cancellation}
\end{figure}

The two panels of \figref{fig:chain-cancellation} test the limits of the A5
composition with a case where the final $z$ boost nearly undoes the first. The
net Lorentz factor is only about $1.5$, but A5 obtains it from intermediate
factors of order $\gamma_z$. The rounding changes sharply between nearby
$\gamma_z$ values, and since the same imperfect composed transform is used for
every one of the $3\times10^5$ events, increasing statistics cannot smooth the
green spikes. Applying its inverse \emph{can} cancel the error at some scan
points, which explains the deep dips in the left panel. In the right panel,
the lab-frame energy calculated before the inverse shows the underlying
composition error more directly. A5 removes intermediate-frame loss for
non-cancelling chains, but not cancellation inside the composition itself.

\begin{figure}[t]
\centering
\includegraphics[width=0.48\textwidth]{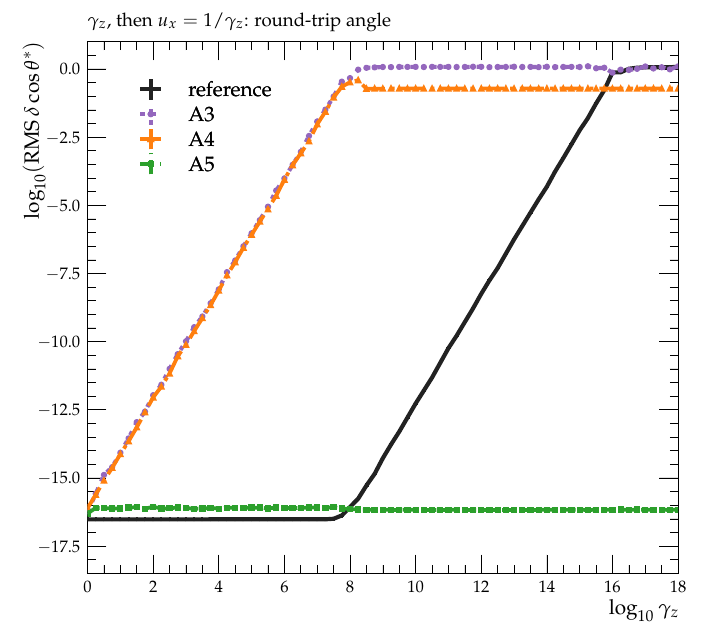}\hfill
\includegraphics[width=0.48\textwidth]{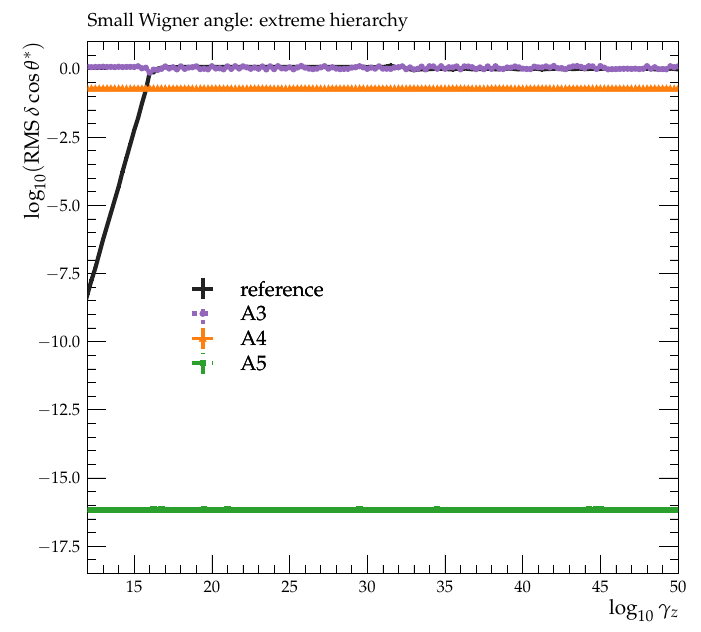}
\caption{The scaled chain $B_x(u_x=1/\gamma_z)B_z(\gamma_z)$, followed by its
inverse, for the same $100~\mathrm{MeV}$ virtual-photon decays as
\figref{fig:chain-order}. The plotted quantity is the RMS error in the reconstructed
rest-frame $\cos\theta^*$. Left: the transition region. Right: the same test
continued to $\gamma_z=10^{50}$. The transverse effect remains finite because
$u_x\gamma_z=1$, while the Wigner angle decreases as $1/\gamma_z$.}
\label{fig:small-wigner}
\end{figure}

The scaled chain in \figref{fig:small-wigner} instead tests a non-cancelling
hierarchy motivated by logarithmic-accuracy studies. After the large $z$
boost, the transverse proper velocity is reduced as $u_x=1/\gamma_z$. A3
constructs and then leaves the highly boosted Cartesian intermediate state,
and loses the rest-frame angle near $\gamma_z=10^8$. A4 protects the first
boost, but the transverse step forces it back through Cartesian components and
gives the same limiting scale. The Cartesian double-double reference postpones
the loss to $\gamma_z\sim10^{16}$.

A5 composes the two boosts first. It applies the resulting Wigner rotation in
the original frame and performs only the final boost in retained light-cone
coordinates. Its round-trip angular error remains at rounding precision
through $\gamma_z=10^{50}$.

\begin{table}[h!]
\centering
\begin{tabular}{lr}
\toprule
Implementation & ns/call\\
\midrule
\csname @@input\endcsname figs/timing.tex
\bottomrule
\end{tabular}
\caption{Direct-call timings at $\gamma=10^6$, reported as the median of seven
passes. ``A4, one shot'' includes conversion from Cartesian to light-cone
coordinates and back; ``A4, retained'' times the rescaling when the vector is
already retained. Values are specific to the compiler and machine.}
\label{tab:timing}
\end{table}

\section{Discussion}
\label{sec:discussion}

The results separate the problem into different parts with different
remedies. A direct-call timing test, with results shown in \autoref{tab:timing}, shows no appreciable cost for retaining the
light-cone representation. Converting to light-cone coordinates and back costs
about the same as the Cartesian formulas; once retained, A4 is two scalar
rescalings. The absolute nanosecond values are compiler- and machine-specific.

\paragraph{Input} A velocity-only interface loses the boost magnitude,
quadratically and then entirely, and no algorithm consuming $\vbeta$ can recover it.
The remedy is to supply $(\gamma,\vu)$, a rapidity, or a four-momentum
\emph{together with} its mass. A four-momentum alone is not enough. As indicated by the survey, real implementations provide interfaces which recompute the mass from $E^2-|\vp|^2$, and thus suffer the output loss on their own input.

\paragraph{Output} A boosted vector stored as Cartesian components does not
contain its own invariants beyond $\gamma\simeq\sqrt{\tol/\varepsilon}$, and the
failure remains invisible over a range $1/\sqrt{\tol}$ above that. Supplying
better input does not help. Where invariants, Minkowski products or rest-frame quantities are
needed at such Lorentz factors, the loss must be prevented by representation, and
where it cannot be prevented, it must be detected.

The nearly null test makes the same point for angles. A cross product is a
useful stable alternative to a dot-product residual, but it still consumes the
representation it is given. Generic rounded directions eventually lose their
small separation. The robust solution is to retain the small transverse
component, angle or invariant when it is known, rather than reconstruct it
afterwards.

\paragraph{Prevention} For boosts along one axis, A4 keeps the vector in light-cone
coordinates aligned with the boost, making the transformation two
scalar rescalings. For chains along different axes, the
chain is composed into one boost and a Wigner rotation before it is applied
(A5), so that the vectors are rotated in a benign frame and boosted once, in
light-cone coordinates. This removes the failure for any chain whose composition is
itself free of cancellation, in particular for perpendicular boosts and for a modest
boost followed by a much larger one. The scaled-chain test shows that this can
remain accurate through hierarchies far beyond the range of a fixed Cartesian
representation. It has three limits. (1) The chain must be known
while the vectors are still in a benign frame, which in practice means carrying
the accumulated transformation with the vectors rather than applying each boost
as it comes. (2) Composition cancels when a later boost nearly undoes an earlier one.
(3) Invariants of vectors with different chains need a common frame, in which the
cancellation can return.

A common example is a boost to a rest frame, an operation there, and the exact
inverse boost. When possible, constructing the operation directly in the original
frame avoids the unstable intermediate representation. Otherwise the transformation
and a stable representation of the intermediate state must be retained.

\paragraph{Detection} Where prevention does not apply, the failure can at least be
made visible. The results show that, below the velocity saturation, nothing in the
output of a Cartesian boost reveals the loss. Results are finite, and the
usual velocity-domain guards act only when $\beta^2$ reaches one. What would reveal it is the
cancellation itself. An operation whose result $r$ is formed from terms $t_i$ has lost
the digits of $\sum_i|t_i|$ that cancel, so a routine could warn or refuse once
$\varepsilon\sum_i|t_i|$ exceeds a chosen fraction $\tol$ of $|r|$. For a boost, the
natural candidates are the boosted energy and the invariant mass $E^2-|\vp|^2$, and
for A5 the composed Lorentz factor. It costs a few operations and needs no change of representation, so it could be added to the boost and mass routines of existing four-vector classes as they are.

\noindent A natural question is why the calculation should not simply be performed in
quadruple precision. Here quadruple precision means the IEEE binary128 format,
with a 113-bit significand; the benchmark reference instead uses a pair of
doubles, with about 106 significant bits and the exponent range of a double.
Both are normally software operations on mainstream processors, and a
binary128 four-vector occupies twice the storage of its double-precision
counterpart. More importantly, promoting an already rounded double input
cannot restore information lost before the call, and this promotion would thus quickly need to propagate throughout the code. Furthermore, higher precision only
moves the invisible-failure window. With $\varepsilon_{128}=2^{-112}$ it starts at
$\gamma\simeq\sqrt{\tol/\varepsilon_{128}}\simeq2\times10^{15}$ for
$\tol=10^{-3}$. The double-double references in
\figref{fig:resonance-extreme} and \figref{fig:small-wigner} exhibit the same
shifted boundary, while the retained representations remain accurate beyond
it.

\section{Code availability}
\label{sec:code}

The accompanying material contains compact C++17 reference implementations of
A1--A5, the three benchmarks, the timing test and the double-double reference.
The benchmarks write YODA objects \cite{Buckley:2023xqh}, and the supplied \texttt{rivet-mkhtml} plot configuration produces the figure panels used here, using a system installation of \rivet. The release also contains the fixed prompt, pinned source
versions and dated output of the static survey from which \autoref{tab:libs},
\autoref{tab:calls} and the implementation findings are taken. The code is also available from \url{https://gitlab.com/bierlich/four-vector}.

\section{Conclusion and practical recommendations}
\label{sec:conclusions}

Lorentz-boost errors can become physically important before any failure is
visible. The source survey
shows that current high-energy physics software overwhelmingly exposes and
stores boosts through Cartesian quantities. The numerical tests establish the
consequence for representative algorithms: reconstructed masses and rest-frame
angles acquire errors of order $\gamma^2\varepsilon$, including when the Lorentz
factor is supplied explicitly, while the usual largest-component accuracy test
still passes. The resonance mass error reaches one per cent at $\gamma=10^7$;
the first non-finite results appear only near $5\times10^7$.

Two losses are responsible. A rounded velocity loses the boost magnitude, and a
boosted vector stored as $(E,p_x,p_y,p_z)$ loses its invariant through
cancellation. More careful arithmetic after either loss cannot recover the
missing information. Special kinematics add algorithmic cancellation, notably
for an antiparallel Cartesian boost and when one transformation nearly undoes
another. Retaining light-cone coordinates along a fixed boost axis avoids the
Cartesian loss through $\gamma=10^{17}$ in the resonance test, without an
appreciable timing penalty. Non-cancelling chains can likewise be composed into
one boost and a Wigner rotation before they are applied; the scaled chain
remains at rounding precision through $\gamma=10^{50}$.

For an author or maintainer of simulation software, the practical conclusions
are:

\begin{itemize}
\item Compare $\gamma^2\varepsilon$ with the required relative tolerance $\tol$ before
choosing an interface or representation. The relevant scale is
$\gamma\simeq\sqrt{\tol/\varepsilon}$, not the much later point at which a
velocity rounds to the speed of light.

\item Pass a boost as $(\gamma,\vu)$, as a rapidity, or as a four-momentum
together with an independently known mass. Do not reconstruct $\gamma$ from a
rounded velocity, and do not expect a compensated evaluation of
$1-\beta^2$ to restore information already lost from that velocity.

\item Keep masses and other known invariants as independent data. Do not
recompute them from $E^2-|\vp|^2$ after a large boost, or use such a recomputed
mass to define another transformation.

\item For nearly collinear directions, a cross product postpones the loss but
does not guarantee accuracy. Retain a known small angular component, angle or
pair invariant instead of reconstructing it from already rounded directions.

\item Where invariants, Minkowski products or rest-frame quantities are needed
after a large boost along a fixed axis, retain the vector in light-cone
coordinates along that axis. Convert back to Cartesian components only when
those components are required.

\item If a chain of boosts is known in advance, accumulate the transformation
and apply one boost and its Wigner rotation while the vectors remain in a
benign frame. Check the composition itself for cancellation: this construction
does not protect a chain in which a later boost nearly undoes an earlier one.
For a boost to a rest frame followed by the inverse boost, formulate the
intermediate operation directly in the original frame when possible; otherwise
retain the transformation and a stable intermediate representation.

\item Add cancellation checks to Cartesian boost, mass and transformation-
composition routines. A finite result and the usual $\beta^2<1$ guard do not
establish accuracy. Warn or refuse when
$\varepsilon\sum_i|t_i|$ exceeds the chosen fraction of a result formed from
the terms $t_i$.

\end{itemize}

Most present collider calculations with modest boosts remain well within the
double-precision range. The same assumption is unsafe for low-mass systems,
auxiliary-frame calculations, single-precision particle records and air-shower
simulations, as well as for the extreme hierarchies used in parton-shower
accuracy tests and ultra-heavy decay calculations. Interfaces which preserve the available information, together
with explicit cancellation checks, make that distinction visible to the
caller instead of leaving an invisible numerical failure.

\section*{Acknowledgements}

I am grateful to Christian Preuss, Torbjörn Sjöstrand and Peter Skands for valuable suggestions and comments on the study and the manuscript. Financial support from Vetenskapsrådet contract 2023-04316 is acknowledged.

\bibliographystyle{elsarticle-num}
\bibliography{bibliography}

\end{document}